\documentclass[dissemination]{jfsma}

\titre{Modélisation à base d'Agents Augmentés par LLM pour les Simulations Sociales :\\ Défis et Opportunités}

\auteur{\"{O}nder G\"{u}rcan}{ongu@norceresearch.no}

\institution{%
  Norwegian Research Center AS (NORCE),
  Universitetsveien 19, Kristiansand, Norway}

\begin{document}

\maketitle

\begin{resume}
Alors que les grands modèles linguistiques (LLM) continuent de faire des progrès significatifs, leur meilleure intégration dans les simulations basées sur des agents offre un potentiel de transformation pour la compréhension des systèmes sociaux complexes.
Cependant, une telle intégration n'est pas triviale et pose de nombreux défis.
Sur la base de cette observation, dans cet article, nous explorons les architectures et les méthodes pour développer systématiquement des simulations sociales augmentées par LLM et discutons des orientations de recherche potentielles dans ce domaine.
Nous concluons que l'intégration des LLM aux simulations basées sur des agents offre un ensemble d'outils puissants pour les chercheurs et les scientifiques, permettant des modèles plus nuancés, réalistes et complets de systèmes complexes et de comportements humains.
\end{resume}

\motscles{Grands modèles de langage (LLM),
Simulations basées sur des agents,
Modélisation des systèmes sociaux,
Orientations de recherche}

\bigskip

\begin{abstract}
As large language models (LLMs) continue to make significant strides, their better integration into agent-based simulations offers a transformational potential for understanding complex social systems. 
However, such integration is not trivial and poses numerous challenges.
Based on this observation, in this paper, we explore architectures and methods to systematically develop LLM-augmented social simulations and discuss potential research directions in this field. 
We conclude that integrating LLMs with agent-based simulations offers a powerful toolset for researchers and scientists, allowing for more nuanced, realistic, and comprehensive models of complex systems and human behaviours. 
\end{abstract}
\keywords{Large Language Models (LLMs),
Agent-Based Simulations,
Social Systems Modeling,
Research Directions}

\section{Introduction}

Les grands modèles linguistiques (LLM) ont connu une expansion rapide dans leur adoption dans une multitude de domaines de recherche et d'applications pratiques ces derniers temps.
Cette prolifération rapide est en grande partie attribuée à leur capacité remarquable à comprendre, générer et traduire le langage humain avec une précision et une fluidité sans précédent.
Des secteurs allant des soins de santé \cite{Clusmann2023,Barrington2023}, où ils aident aux soins aux patients et à la recherche médicale, à la finance \cite{Wu2023}, pour analyser les tendances du marché et automatiser le service client, ont exploité les capacités des LLM pour améliorer l'efficacité et l'innovation.
De plus, dans le domaine universitaire, ces modèles jouent un rôle essentiel dans l'analyse de vastes ensembles de données pour obtenir des informations \cite{Taylor2022}, accélérant ainsi les résultats de la recherche dans des domaines tels que les sciences sociales, la linguistique et l'informatique.
La polyvalence et la sophistication évolutive des LLM ont ainsi consolidé leur rôle de technologie fondamentale qui remodèle le paysage de l'industrie et de la recherche, favorisant de nouvelles méthodologies et approches dans toutes les disciplines.

En sciences sociales, les simulations sociales sont un domaine dans lequel les LLM peuvent être efficacement appliqués.
Les simulations sociales sont utilisées pour modéliser et analyser les interactions complexes au sein des systèmes sociaux, y compris des facteurs tels que les comportements individuels, la dynamique de groupe, les normes sociales et les structures institutionnelles.
Ces simulations visent à comprendre, prédire ou examiner des scénarios hypothétiques au sein des systèmes sociaux.
En ce sens, une technique utilisée pour les simulations sociales est la modélisation basée sur les agents (ABM).
L'ABM est utilisée pour concevoir et simuler les actions et les interactions d'agents autonomes (individus, entités ou organisations) afin d'évaluer leurs effets sur le système dans son ensemble.

L'ABM peut utiliser des techniques d'IA pour améliorer la complexité, l'adaptabilité et le réalisme des modèles.
Jusqu'à présent, l'ABM a déjà intégré de manière pratique des techniques d'IA telles que l'apprentissage automatique (ML) \cite{Turgut2023,Dehkordi2023,Chen2021,Vahdati2019}, l'apprentissage par renforcement \cite{Chmura2007,Chen2018} et l'apprentissage par renforcement inverse \cite{Lee2018} dans diverses études de simulation sociale.
Cependant, le potentiel des LLM pour aider à la compréhension de systèmes sociaux complexes n'a pas encore été exploité.
Il existe peu d'études axées sur cette intégration récemment \cite{Rask2024,Shen2023}.
Même si ces études fournissent des solutions pratiques, elles manquent d'une base conceptuelle bien définie qui est nécessaire pour explorer des architectures et des méthodologies à usage général - en étendant éventuellement celles existantes - qui sont efficaces pour intégrer de manière transparente et systématique les simulations sociales et les LLM.

En outre, les LLM peuvent rationaliser et améliorer divers autres aspects du processus ABM, tels que l'analyse de la littérature, la préparation et l'interprétation des données, l'étalonnage des paramètres, les analyses de sensibilité et les analyses des résultats.
Nous avons déjà commencé à voir certains impacts des LLM sur ces activités.
Mais ils sont rares et ne fournissent pas une vision claire en termes de simulations sociales.
En nous appuyant sur ces observations, nous présentons dans cet article des pistes de recherche prospectives sur l'augmentation des simulations basées sur des agents grâce à l'intégration des LLM.
Nous articulons une exploration structurée du potentiel symbiotique entre les LLM et les cadres ABM, visant à faire progresser les fondements méthodologiques et à améliorer les capacités analytiques des simulations sociales.

Le reste de l'article est organisé comme suit.
Nous fournissons d'abord un aperçu des LLM et ABM pour les simulations sociales dans les sections \ref{sec:Large-Language-Models} and \ref{sec:ABM-in-Social-Simulations}. 
Ensuite, nous identifions la base conceptuelle en évaluant diverses méthodologies pour la construction de systèmes multi-agents dans la section \ref{sec:The-Conceptual-Baseline}.
Ensuite, nous fournissons un aperçu de certaines des principales directions de recherche utiles pour développer l'idée dans la section \ref{sec:Research-Directions} et enfin dans la section \ref{sec:Conclusions}, nous concluons l'article.

\section{LLMs}
\label{sec:Large-Language-Models}

Un grand modèle de langage (LLM) peut être défini comme une fonction qui recherche, en considérant une série de tokens (tels que des mots, des fragments de mots, des signes de ponctuation, des émojis, etc.), quels tokens sont les plus susceptibles de suivre ensuite.
Aujourd'hui, de nombreux LLM sont accessibles au public.
Gemini\footnote{\url{https://gemini.google.com/}, dernier accès le 26/08/2024.} de Google, plus efficace et compact avec jusqu'a 540 milliards de paramètres, est accessible gratuitement via un navigateur Web et une API, avec un ensemble des données d'entraînement diversifié \cite{Geminiteam2024}.
Le LLaMA\footnote{\url{https://llama.meta.com}, dernier accès le 28/06/2024.} de Meta, destiné à l'avancement de la recherche en IA, propose divers modèles avec jusqu'à 70 milliards de paramètres accessibles aux chercheurs via l'application \cite{Touvron2023}.
Les modèles GPT d'OpenAI utilisent l'architecture de transformateur pour la génération de sortie dynamique \cite{Vaswani2017}, avec différentes versions accessibles différemment : GPT-3.5 est gratuit via une interface Web\footnote{\url{https://chat.openai.com}, dernier accès le 28/06/2024.}, tandis que GPT-4 nécessite un abonnement, l'utilisation de l'API étant également payante à l'utilisation en fonction de la tokenisation pour le traitement du langage naturel.
Les API LLM sont des interfaces qui permettent aux développeurs d'accéder aux capacités de ces LLM avancés dans leurs applications.
Les API peuvent être intégrées dans des applications utilisant n'importe quel langage de programmation capable d'effectuer des requêtes HTTP, généralement en envoyant l'invite au LLM à traiter avec divers paramètres qui ajustent le comportement du LLM (comme la version du modèle de langage à utiliser et la température qui contrôle le caractère aléatoire de la sortie générée).

Un LLM peut être considéré comme un simulateur non déterministe avec la capacité de jouer un nombre infini de personnages.
Essentiellement, les LLM peuvent être affinés en les exposant à des rôles spécifiques afin qu'ils puissent simuler des interactions de type humain.
Le réglage fin est obtenu en entraînant le modèle sur un ensemble de données organisé qui incarne le langage, les connaissances et les nuances des rôles qu'il est censé jouer.
Ce processus implique d'ajuster les paramètres du modèle afin qu'il s'aligne mieux sur les modèles, le vocabulaire et les processus de prise de décision caractéristiques des rôles cibles.
Lors du réglage fin, le modèle apprend à prioriser les réponses qui reflètent les traits spécifiques, l'expertise ou la personnalité des rôles en question.
Cela se fait souvent en utilisant un ensemble de données plus petit et plus spécialisé après que le modèle a été pré-entraîné sur un large corpus de texte, ce qui lui permet d'adapter ses vastes connaissances générales à des contextes et comportements plus étroitement définis.
Cependant, un tel ensemble de données peut également être relativement volumineux.

Une architecture bien connue pour la création de systèmes qui exploitent les LLM sur des ensembles de données spécialisés relativement volumineux est appelée Retrieval Augmented Generation (RAG) \cite{Lewis2021}.
Dans cette architecture, l'ensemble de données est divisé en morceaux (comme quelques paragraphes ou une page), puis ces morceaux sont envoyés à un LLM et transformés en vecteur.
Chaque morceau aura un vecteur (c'est-à-dire une série de nombres) qui est une représentation numérique de l'essence de ce morceau.
Après chaque fois qu'une invite est envoyée, son vecteur est également calculé à l'aide du même LLM.
Ensuite, les morceaux les plus proches sont trouvés en effectuant une comparaison mathématique entre les vecteurs de morceaux et le vecteur d'invite.
Enfin, ces morceaux sont utilisés dans le cadre de l'invite.

\section{ABM en Simulations Sociales}
\label{sec:ABM-in-Social-Simulations}

L'ABM est une technique essentielle dans l'exploration et la compréhension des systèmes sociaux par le biais de simulations informatiques \cite{Macal2016}.
Les systèmes sociaux, dans ce contexte, font référence à des réseaux complexes d'interactions entre les individus, les institutions et leurs environnements (qui peuvent inclure des facteurs tels que les comportements individuels, la dynamique de groupe, les normes sociales et les structures institutionnelles).
La nature multiforme des systèmes sociaux permet aux individus de jouer plusieurs rôles simultanément (comme parent, employé, consommateur et citoyen dans une société humaine), chacun avec son propre ensemble d'attentes et de normes.

L'ABM pour les systèmes sociaux vise à imiter les processus sociaux en simulant les actions et les interactions des agents, qui représentent des individus ou des entités au sein de ces systèmes, afin de prédire et de comprendre des phénomènes complexes.
En d'autres termes, l'ABM améliore l'étude des systèmes sociaux en offrant une approche de modélisation ascendante, où les phénomènes d'intérêt au niveau macro émergent des interactions au niveau micro des agents.
Cette capacité est inestimable en sciences sociales, où la compréhension de l'émergence de phénomènes sociaux complexes à partir de simples règles d'interaction peut fournir des informations approfondies sur la nature de l'ordre social, l'évolution des normes et des institutions et la dynamique du changement social.
Les simulations sociales permettent d'examiner des scénarios hypothétiques, de tester des théories du comportement et d'explorer les effets des décisions politiques sans les contraintes des expériences réelles.

Cependant, l’analyse des résultats des  ABM pour comprendre les systèmes sociaux présente plusieurs défis, principalement en raison de la complexité et du dynamisme inhérents aux modèles et aux systèmes qu’ils cherchent à représenter.
Tout d’abord, les ABMs génèrent souvent de grandes quantités de données par la simulation d’interactions entre de nombreux agents au fil du temps, ce qui rend difficile de discerner des modèles clairs ou de tirer des conclusions simples.
Les phénomènes émergents, une caractéristique des ABMs, bien que précieux pour comprendre les résultats au niveau macro des comportements au niveau micro, peuvent compliquer l’analyse car ces résultats ne sont pas toujours prévisibles ou linéaires.
Deuxièmement, interpréter les résultats des ABMs d’une manière qui soit significative pour l’élaboration des politiques ou les avancées théoriques, nécessite de combler le fossé entre les résultats complexes, souvent techniques, des modèles et les cadres conceptuels des sciences sociales.
Cela exige souvent une collaboration interdisciplinaire pour garantir que les informations générées sont à la fois rigoureuses sur le plan scientifique et pertinentes sur le plan social.

En plus de l'ABM, les chercheurs en sciences sociales utilisent diverses autres méthodes pour étudier les systèmes sociaux.
Des enquêtes et des questionnaires sont utilisés pour recueillir des données à grande échelle sur les attitudes, les comportements et les expériences des individus.
Les entretiens et l'ethnographie offrent des aperçus qualitatifs approfondis des phénomènes sociaux, saisissant les nuances du comportement humain et des interactions sociales.
Les études de cas fournissent une analyse détaillée d'instances ou d'événements spécifiques, permettant une compréhension approfondie des processus sociaux complexes.
Des modèles expérimentaux et quasi expérimentaux sont utilisés pour établir des relations causales entre les variables.
L'analyse statistique et les techniques informatiques, notamment l'analyse de réseau et l'exploration de données, sont appliquées pour analyser et modéliser de grands ensembles de données.
Enfin, l'analyse du contenu et du discours est utilisée pour examiner les modèles de communication et la construction du sens dans les contextes sociaux.
Certaines de ces méthodes sont déjà utilisées dans les études ABM.

\section{La Base Conceptuelle}
\label{sec:The-Conceptual-Baseline}

L'intégration des LLM dans les ABM pour les simulations sociales offre un potentiel de transformation pour la compréhension des systèmes sociaux complexes.
Cependant, une telle intégration nécessite une base conceptuelle pour relier les deux domaines.
Une base conceptuelle est un cadre fondamental clair et cohérent qui décrit les concepts, variables, hypothèses et relations clés au sein d'un système ou d'un modèle donné.
Elle sert de point de référence pour comprendre la dynamique et le comportement du système étudié, et elle est ensuite utilisée pour le développement et l'analyse du modèle.
Une telle base conceptuelle peut être établie à l'aide des méthodologies d'ingénierie pour les  systèmes multi-agents (SMA).
Un SMA est généralement considéré sous quatre angles principaux : agent, interaction, environnement et organisation.

Les méthodologies orientées agents donnent la priorité aux agents individuels, à leur autonomie, à leurs états internes et à leur prise de décision \cite{Shoham1993,Cuesta2008,Abdalla2021}. 
Elles utilisent le modèle des croyances, des désirs et des intentions (BDI) pour la cognition des agents et définissent les capacités et les rôles des agents. Cette approche permet une modélisation détaillée d'agents complexes et autonomes qui prennent des décisions basées sur des perceptions et remplissent des rôles dans les interactions multi-agents.
Les méthodologies orientées interaction se concentrent sur la communication des agents et la dynamique de coordination, en utilisant des protocoles et des messages pour définir des modèles d'interaction \cite{Chopra2023}. 
Les protocoles assurent des échanges d'informations structurés, tandis que les mécanismes de négociation et de coordination soutiennent la prise de décision conjointe et la synchronisation des actions. Les normes et conventions sociales établissent un cadre pour des interactions prévisibles et le respect de règles de conduite partagées.
Les méthodologies orientées vers l'environnement mettent l'accent sur le rôle de l'environnement dans les interactions entre agents, en se concentrant sur les ressources partagées, la stigmergie pour la coordination indirecte et les affordances qui dictent les actions des agents en fonction de leurs capacités. \cite{Ricci2011,Weyns2006}.
Les méthodologies orientées vers l'organisation résument le MAS en ce qui concerne les groupes, les équipes et les structures organisationnelles plus larges. \cite{Abbas2015,Criado2013,Giorgini2006,Dignum2005,Ferber2004,Hubner2002,Roussille2022}.
Ces méthodologies se concentrent sur la manière dont les agents sont organisés et interagissent au sein d'entités plus vastes, définissant les rôles, les responsabilités et les relations à travers les groupes et les équipes.
Les structures organisationnelles, telles que les hiérarchies ou les réseaux, dictent le flux d'informations et le contrôle entre les agents.
Les politiques et les réglementations établissent les règles régissant le comportement au sein de l'organisation, guidant les actions et les interactions des agents.

L'approche MAS orientée vers l'organisation est particulièrement adaptée à la modélisation des systèmes sociaux et à l'intégration des LLM en raison de l'accent mis sur les interactions et les rôles structurés au sein d'un système complexe.
Cette approche reflète la nature hiérarchique et en réseau des systèmes sociaux, où les entités (agents) assument des rôles et des responsabilités spécifiques régis par des normes et des politiques établies.
Une telle structure organisationnelle permet de définir clairement les rôles des LLM, facilitant leur intégration en tant qu'agents avec des fonctions spécifiques liées à la compréhension, à la génération et au traitement du langage. Cela améliore non seulement la capacité du système à imiter les structures sociales humaines, mais exploite également les capacités des LLM en matière de traitement du langage naturel pour améliorer la communication et la coordination entre les agents.
En alignant les aspects structurels et fonctionnels des MAS avec les propriétés inhérentes aux systèmes sociaux et les points forts des LLM, l'approche orientée vers l'organisation offre un cadre robuste pour capturer la complexité et la dynamique des interactions sociales, ce qui en fait un choix supérieur pour ces applications.
Par conséquent, nous affirmons qu'un outil ABM efficace augmenté par LLM devrait soutenir la base conceptuelle orientée vers l'organisation.

Dans cette optique, nous proposons de définir les agents dans les simulations sociales comme des agents sociaux qui jouent le rôle d'un ou de plusieurs personnages prédéfinis \cite{Shanahan2023,Andreas2022,Park2023}.
Les compétences d'un agent social sont les capacités de jeu de rôle qu'il possède grâce à ses interactions avec les environnements, et sont utilisées dans une communauté d'autres acteurs qui habitent également ces environnements.

\section{Les Directions de la Recherche}
\label{sec:Research-Directions}

Dans cette section, nous explorons les principales pistes de recherche qui semblent pertinentes pour transformer les simulations sociales à l’aide des LLM.

\subsection{Les Revues Littéraires}

Le volume de la littérature scientifique est écrasant \cite{Hutson2020,Wagner2022}, avec des niveaux de qualité variables selon les publications.
Il est nécessaire de disposer d'outils capables de rechercher, d'évaluer et de résumer la littérature scientifique de manière objective et efficace.
Les LLM peuvent relever de manière significative les défis du processus d'examen traditionnel grâce à leurs capacités avancées de traitement efficace de vastes quantités de données textuelles \cite{Wagner2022,Dunn2022,Zhang2023b} et ils sont moins susceptibles de sélectionner la littérature pour étayer les hypothèses souhaitées (c'est-à-dire réduire le biais du chercheur) \cite{Muller2022}.
En automatisant le filtrage et le résumé initiaux de la littérature, les LLM peuvent aider les chercheurs à résoudre le problème de la surcharge d'informations, leur permettant d'identifier rapidement les études pertinentes sans compromettre l'étendue ou la profondeur de l'examen.
Leur capacité à analyser et à résumer des textes dans plusieurs langues peut également surmonter les barrières linguistiques, offrant ainsi l'accès à un éventail plus large de littérature.

\subsection{Modélisation d'Architectures}

Comme indiqué dans la section \ref{sec:The-Conceptual-Baseline}, l'approche orientée organisation qui définit les agents comme des acteurs jouant un rôle fournit une base solide pour la modélisation des agents sociaux.
%
%
Cependant, il existe plusieurs architectures orientées organisation et l'adéquation aux simulations sociales augmentées par LLM reste à étudier.
De plus, des recherches devraient également être menées sur la conception et la réutilisation efficaces des rôles des agents sociaux.
Une conception efficace des rôles est nécessaire pour obtenir des informations efficaces (voir la section \ref{sec:Obtaining-Insights}) et une réutilisation efficace des rôles est nécessaire pour des simulations sociales à grande échelle et reproductibles.
De plus, les LLM peuvent être utilisés pour générer des modèles et des scénarios basés sur des agents en utilisant le langage naturel et les données qualitatives associées.

\subsection{Préparation des Données}
\label{sec:Data-Preparation}

La collecte de données pour les simulations sociales prend du temps et coûte cher car elle fait face à plusieurs défis clés, notamment la capture de la complexité des systèmes sociaux, la garantie de données de haute qualité et pertinentes, la prise en compte des préoccupations éthiques et de confidentialité, l'intégration de diverses sources de données et la réalisation d'une modélisation et d'une simulation précises.
De plus, l'étalonnage et la validation des modèles, la gestion des contraintes informatiques, la promotion de la collaboration interdisciplinaire et l'adaptation aux systèmes sociaux dynamiques ajoutent des niveaux de complexité.
Enfin, assurer la généralisabilité et la transférabilité des modèles à différents contextes ou populations est un défi important, nécessitant une approche prudente et méthodique et souvent une collaboration interdisciplinaire.

Les LLM ont le potentiel d'améliorer considérablement le processus de collecte de données pour les simulations sociales en offrant des solutions à de nombreux défis décrits.
Pour capturer la complexité des systèmes sociaux, les LLM peuvent traiter et analyser de grands volumes de données textuelles provenant de diverses sources \cite{Zhang2023,Semeler2024}, offrant une compréhension nuancée de la dynamique sociale qui peut éclairer des modèles plus précis et plus complets.
Leurs capacités avancées de traitement du langage naturel permettent l’intégration de types de données variés, des données structurées au texte non structuré, facilitant la création d’ensembles de données multidimensionnels plus riches.

\subsection{Dataficiation}
\label{sec:Dataficiation}

La datafication est la transformation d'interactions et de phénomènes sociaux complexes en données quantifiables, permettant un suivi en temps réel et une analyse prédictive \cite{MayerSchnberger2014}.
Les agents sociaux augmentés par LLM peuvent jouer un rôle central dans le processus de datafication.
Ces données peuvent ensuite être analysées pour découvrir des modèles, des tendances et des informations sur les comportements et les systèmes sociaux.
Par exemple, les agents sociaux interagissent dans des simulations, générant continuellement des données reflétant les changements dans le temps, y compris l’évolution des systèmes sociaux face aux pressions externes ou dynamiques internes. Cet aspect dynamique des données est crucial pour étudier les processus de changement social, la diffusion de l’innovation et l’émergence des normes sociales.

\subsection{Obtenir des Informations}
\label{sec:Obtaining-Insights}

Les simulations sociales produisent souvent des données trop volumineuses et trop complexes pour être conservées et analysées.
Il est démontré qu'en utilisant les LLM, il est possible d'obtenir des informations à partir des données.
Il est donc possible d'obtenir des informations sur les agents sociaux en entamant simplement un dialogue avec eux.
Si elle est préparée correctement, il est possible de simuler une population synthétique d'une vaste gamme d'agents sociaux représentant les expériences et les perspectives humaines \cite{Fuchs2023}, ce qui peut fournir une représentation plus précise du comportement humain et de la dynamique sociale que ce qui est réalisable par les méthodes traditionnelles \cite{Grossmann2023}. 
Avec un conditionnement approprié \cite{Argyle2023}, les agents sociaux peuvent jouer le rôle de personnages qui ont des croyances et des intentions et qui fournissent des réponses précises et objectives aux questions des utilisateurs \cite{Mei2024}. 
Ils sont idéaux pour obtenir des informations car ils "peuvent répondre rapidement à des centaines de questions sans se fatiguer" et "ont besoin de moins d'incitations que les humains pour donner des réponses fiables" \cite{Dillion2023}.
Des études récentes montrent que les LLM peuvent avoir la capacité de faire des jugements assez bien alignés sur les jugements humains \cite{Dillion2023}.
Les informations recueillies lors des différents dialogues peuvent ensuite être organisées sous forme de données quantifiables pouvant faire l'objet d'une analyse statistique afin de donner un aperçu des tendances et des schémas sociaux plus larges.
Une telle approche peut alors être utilisée pour générer des hypothèses et les valider dans les sociétés humaines \cite{Argyle2023,Park2023}.

\subsection{Explicabilité}

Les agents sociaux augmentés par LLM peuvent générer des explications en langage naturel pour leurs actions, leurs décisions et les mécanismes sous-jacents de la simulation \cite{Peterson2021,Gil2017}.
Cela peut rendre le comportement de ces agents plus compréhensible pour les chercheurs (de diverses disciplines, et pas seulement pour ceux qui ont une formation en informatique), les parties prenantes et le grand public, en traduisant des algorithmes et des processus décisionnels complexes en explications faciles à assimiler.
En outre, en tirant parti des vastes connaissances et de la compréhension des dynamiques sociales intégrées dans les LLM, ces agents augmentés peuvent fournir des informations contextuelles sur leur comportement. 
Par exemple, ils peuvent expliquer comment certaines normes sociales, certains événements historiques ou certains aspects culturels influencent leurs actions dans la simulation, ce qui permet de mieux comprendre les phénomènes sociaux modélisés.

\subsection{Plateformes et outils}

Pour être efficaces, les simulations sociales augmentées par les LLM nécessitent des outils sophistiqués, conçus pour gérer les complexités des interactions sociales dynamiques et des processus de classification des données. 
Ces outils doivent faciliter l'intégration transparente des LLM dans les plates-formes de simulation, en offrant des caractéristiques pour une configuration facile, des ajustements en temps réel et un traitement éthique des données. 
Il est important que la conception de ces outils soit fondée sur une base conceptuelle orientée vers l'organisation (voir la section \ref{sec:The-Conceptual-Baseline}), garantissant qu'ils s'alignent sur la dynamique structurée des systèmes sociaux et qu'ils soutiennent une recherche ciblée et interdisciplinaire. 
Cette approche améliore l'accessibilité, la précision et la conformité éthique des simulations, permettant aux chercheurs d'explorer et de comprendre en profondeur les phénomènes sociaux.

\section{Conclusions}
\label{sec:Conclusions}

Les LLM offrent un cadre transformateur pour la simulation et l'analyse des systèmes sociaux.
Dans cette étude, nous avons proposé différentes interventions LLM qui couvrent l'ensemble de la chaîne de simulation sociale.
En plus de raisonner le comportement des agents impliqués, les LLM peuvent permettre des interactions plus prospères et intuitives entre les utilisateurs et les agents simulés.
Cela peut rendre la simulation sociale plus accessible et conviviale, permettant aux individus de diverses disciplines, y compris les sciences sociales, les soins de santé, la planification urbaine et les études environnementales, de les utiliser.
Cette démocratisation peut permettre à ces professionnels de modéliser et d'analyser des systèmes complexes pertinents pour leur domaine sans avoir besoin d'une expertise approfondie en informatique.
En conséquence, nous pouvons nous attendre à une collaboration accrue entre les informaticiens et les experts d'autres domaines, ce qui conduira à des approches plus interdisciplinaires et à des solutions innovantes aux problèmes complexes du monde réel.

Les avantages potentiels des grands modèles de langage (LLM) méritent d'être sérieusement pris en compte. 
Cependant, il est essentiel que les scientifiques et les développeurs travaillant avec des outils de modélisation basés sur des agents (ABM) améliorés par les LLM envisagent également la possibilité que ces outils puissent, dans certaines circonstances, entraver la connaissance scientifique au lieu de la faire progresser (e.g., en cas d'hallucinations \cite{Ji2023}). 
Cela signifie que si les LLM offrent des avantages épistémiques significatifs, ils posent également des risques épistémiques si les scientifiques s'appuient sur eux en tant que partenaires dans la production de connaissances \cite{Messeri2024}.
En effet, traiter les outils ABM augmentés de LLM comme des collaborateurs dans la recherche scientifique expose les scientifiques au risque de tomber dans des illusions de compréhension, qui est une classe d'erreur métacognitive qui se produit lorsque les individus ont des croyances erronées sur la profondeur ou la précision de leur propre compréhension \cite{Rozenblit2002,Rabb2019}.

\section*{Remerciements}

Le travail présenté ici fait partie du projet URBANE, qui a reçu un financement de l'action d'innovation Horizon Europe de l'Union européenne dans le cadre de la convention de subvention no 101069782.



\bibliographystyle{plain} 
{\small \bibliography{references}}

\end{document}